\documentclass[12pt]{article}

\usepackage{scicite}

\usepackage{times}

\usepackage[labelfont=bf,labelsep=period]{caption}

\usepackage{amsmath,mathtools}
\usepackage{hyperref,graphicx,multirow,multicol,color}

\newenvironment{sciabstract}{%
\begin{quote} \bf}
{\end{quote}}

\newcounter{lastnote}
\newenvironment{scilastnote}{%
\setcounter{lastnote}{\value{enumiv}}%
\addtocounter{lastnote}{+1}%
\begin{list}%
{\arabic{lastnote}.}
{\setlength{\leftmargin}{.22in}}
{\setlength{\labelsep}{.5em}}}
{\end{list}}

\title{Detection of the Aromatic Molecule Benzonitrile ($c$-C$_6$H$_5$CN) in the Interstellar Medium}

\author
{Brett A. McGuire,$^{1,2,\ast}$ Andrew M. Burkhardt,$^{3}$ Sergei Kalenskii,$^4$\\ Christopher N. Shingledecker,$^5$ Anthony J. Remijan,$^1$  Eric Herbst,$^{5,3}$\\ and Michael C. McCarthy$^{2,6}$\\
\\
\normalsize{$^{1}$National Radio Astronomy Observatory, Charlottesville, VA 22903, USA}\\
\normalsize{$^{2}$Harvard-Smithsonian Center for Astrophysics, Cambridge, MA 02138, USA}\\
\normalsize{$^{3}$Department of Astronomy, University of Virginia, Charlottesville, VA 22904 USA}\\
\normalsize{$^{4}$Astro Space Center, Lebedev Physical Institute, Russian Academy of Sciences, Moscow, Russia}\\
\normalsize{$^{5}$Department of Chemistry, University of Virginia, Charlottesville, VA 22904 USA}\\
\normalsize{$^{6}$School of Engineering and Applied Sciences, Harvard University, Cambridge, Massachusetts 02138 USA}\\
\\
\normalsize{$^\ast$To whom correspondence should be addressed; E-mail: bmcguire@nrao.edu.}
}

\date{}

\begin{document} 


\baselineskip24pt


\maketitle


\begin{sciabstract}

Polycyclic aromatic hydrocarbons and polycyclic aromatic nitrogen heterocycles are thought to be widespread throughout the Universe, because these classes of molecules are probably responsible for the unidentified infrared bands, a set of emission features seen in numerous Galactic and extragalactic sources.  Despite their expected ubiquity, astronomical identification of specific aromatic molecules has proven elusive.  We present the discovery of benzonitrile ($c$-C$_6$H$_5$CN), one of the simplest nitrogen-bearing aromatic molecules, in the interstellar medium.  We observed hyperfine-resolved transitions of benzonitrile in emission from the molecular cloud TMC-1. Simple aromatic molecules such as  benzonitrile may be precursors for polycyclic aromatic hydrocarbon formation, providing a chemical link to the carriers of the unidentified infrared bands.

\end{sciabstract}

The mid-infrared spectra, roughly from 3 to 20 $\mu$m, of the interstellar medium (ISM) and photodissociation regions (PDRs) in both our Galaxy \cite{Low:1984ld} and external galaxies \cite{Regan:2004cb} are dominated by emission features commonly referred to as the unidentified infrared (UIR) bands.  Due to the close agreement of prominent UIR features with the characteristic vibrational frequencies of aromatic C-C and C-H bonds, it is now widely accepted that polycyclic aromatic hydrocarbons (PAHs), and fullerenes like the recently detected C$_{60}^{\hspace{0.25em}+}$ molecule \cite{Campbell:2015hp}, and their closely related derivatives, are probably the carriers responsible for most of these features \cite{Tielens:2008fx}.  A substantial fraction of interstellar carbon is calculated to be in the form of PAHs ($\sim$10\%; \cite{Chiar:2013bj}), yet the origin of these aromatic species is a topic of considerable debate.  In the diffuse ISM and PDR regions, where 30--60\% of the carbon is locked up in dust grains \cite{Draine:2016jy}, top-down models of PAH formation --  through the destruction of dust grains by the harsh radiation environment, shock waves, or both -- may be viable pathways \cite{Berne:2015fi}.  In denser molecular clouds which are not subject to the ultraviolet radiation and which haven't been subject to shocks, other pathways must exist to synthesize these species from smaller precursor molecules. 

Despite the widespread acceptance of PAHs as a common class of interstellar molecules, no specific PAH has been identified in the ISM, either by rotational spectroscopy or by observations of its infrared features, despite long and sustained efforts \cite{KOFMAN201719}.    In the microwave and (sub-)millimeter regimes, while some laboratory data do exist \cite{Thorwirth:2007ic}, such studies are relatively uncommon.  Many PAHs are poor candidates for detection through radio astronomy, both because of unfavorably large rotational partition functions and because they are either apolar or weakly polar, and thus lack sufficiently intense rotational lines (compared to linear molecules of similar composition and size).  A notable exception is corrannulene (C$_{20}$H$_{10}$), a bowl-shaped molecule with a relatively large permanent dipole moment (2.07~Debye (D); \cite{Lovas:2005jn}), but astronomical searches for that molecule have been unsuccessful as well \cite{Pilleri:2009jy}.  In the infrared, while a concerted effort has been undertaken to catalog both laboratory and theoretical vibrational and Raman spectra of PAHs \cite{Boersma:2014bb}, the structural similarities among individual species result in spectra that are often indistinguishable at the modest resolving power that can routinely be achieved by astronomical observations; as a result, aggregate spectra consisting of many PAHs are invoked to reproduce astronomical features \cite{Tielens:2008fx}.

\begin{figure}[t]
\centering
\includegraphics[width=\textwidth]{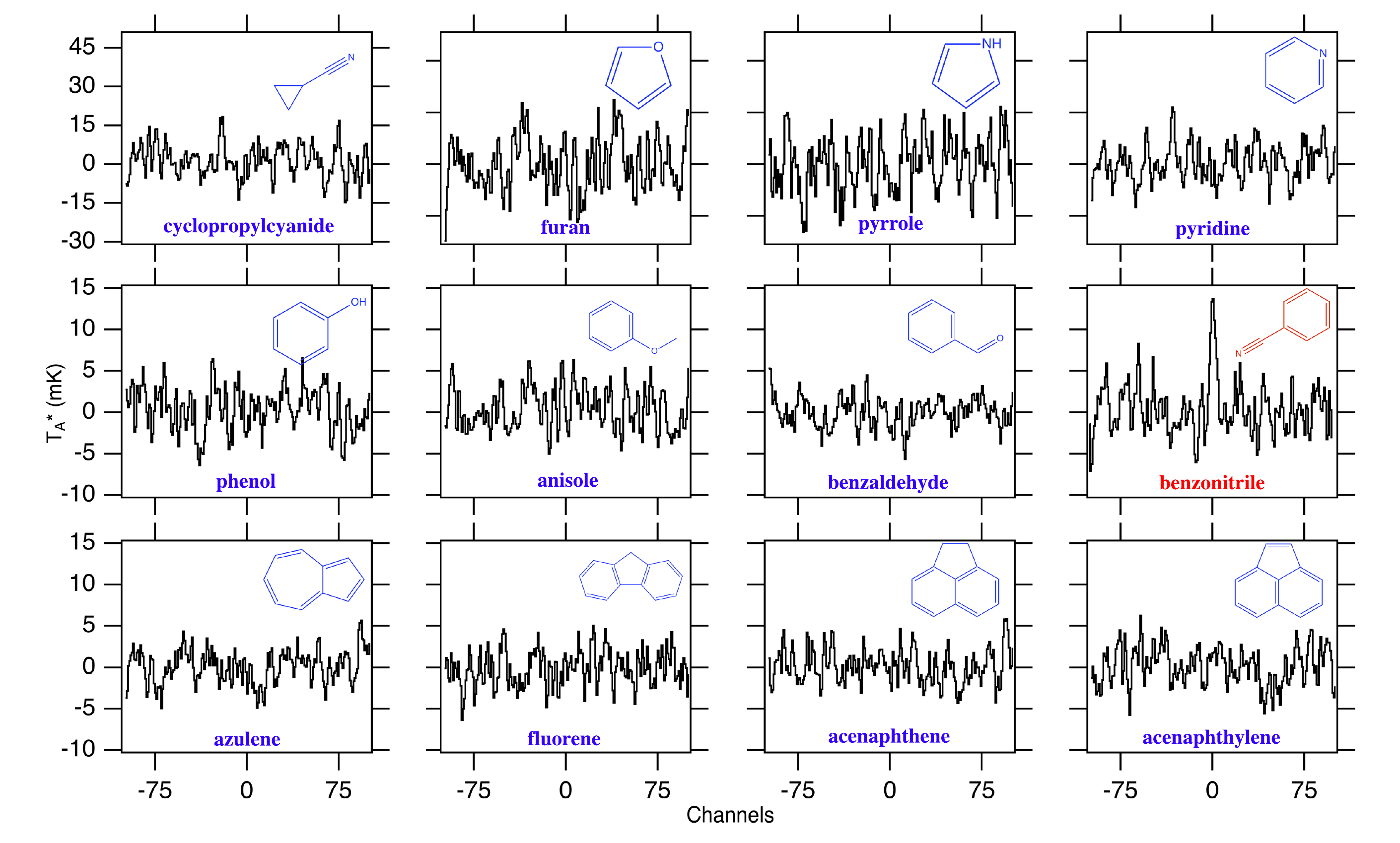}
\caption{\textbf{Composite averages of molecules toward TMC-1.} Velocity-stacked composite averages of all transitions of a given molecule with upper state energy ($E_U$)~$<$~70~K constructed from the entire survey (8.8--50~GHz) of TMC-1 \cite{Kaifu:2004tk}.  The channel spacing is 20~kHz. If a molecule is present, signal in antenna temperature (T$_{\rm{A}}^*$) would be expected at channel 0. A detectable signal was only present for benzonitrile (red).}
\label{cas}
\end{figure}

For these reasons, many attempts to understand the chemistry of PAHs have focused on the possible formation pathways which proceed through more readily detectable molecules.  Much effort has been centered on modeling the formation of small five- and six-membered aromatic rings, and their subsequent reactions with smaller hydrocarbons and nitrogen species to produce PAHs and polycyclic aromatic nitrogen heterocycles (PANHs; \cite{Kaiser:2015ef}).  To date, the only interstellar detection of a five or six-membered aromatic ring is benzene (C$_6$H$_6$), through the observation of a single weak absorption feature arising from its $\nu_4$ bending mode near 14.85 $\mu$m in a handful of sources \cite{Cernicharo:2001mw,Malek:2011kj,Kraemer:2006va,GarciaHernandez:2016tn}.  The lack of a permanent dipole moment, however, precludes the identification of benzene via its rotational transitions.

In this study, we searched for a number of simple aromatic molecules, including several PA(N)Hs and nitriles (R-C$\equiv$N), a class of molecules believed to give rise to a common UIR feature at 6.2 $\mu$m \cite{Hudgins:2005nq}. The astronomical source targeted in these observations was the cold core Taurus Molecular Cloud 1 (TMC-1), which has long been known to display a rich chemistry dominated by unsaturated carbon-chain molecules such as the cyanopolyynes (HC$_n$N; $n$~=~odd) (e.g. \cite{McGuire:2017ud,Loomis:2016jsa,Kaifu:2004tk,Gratier:2016fj}).  The initial search was performed by construction of velocity-stacked composite-average spectra of twelve target molecules (Figure~\ref{cas}; \cite{Kalenskii:2017ta}) using existing survey data taken with the Nobeyama 45 m telescope \cite{Kaifu:2004tk}.  This method enhances the signal-to-noise ratio (SNR) of a potential molecular detection by averaging the signal from multiple transitions of a molecule in velocity space.   These composite averages are effective preliminary indicators of a molecule in a source such as TMC-1, where spectral features are narrow (0.3--0.5~km~s$^{-1}$), the line density is relatively low ($\sim$1~line per 200~km~s$^{-1}$), and the molecules occupy a narrow range in local standard of rest (LSR) velocity ($v_{\rm{lsr}}$~=~5.5--5.9~km~s$^{-1}$) \cite{Kaifu:2004tk}.  As shown in Figure~\ref{cas}, the composite spectra show highly suggestive evidence for benzonitrile in this source.    Nevertheless, the observation of individual transitions is required to establish a firm detection, and to enable the robust determination of the molecular abundance.  The sensitivity and spectral resolution of the existing survey observations, however, were insufficient for that task.

We performed observations with the 100 m Robert C. Byrd Green Bank Telescope to confirm the detection of benzonitrile, by observing nine of its individual rotational transitions, using deep integrations at high spectral resolution.  Because the spectral features of other molecules in TMC-1 are so narrow, $^{14}$N nuclear hyperfine structure is expected to be partially resolved in benzonitrile's lower rotational transitions.  Existing spectral catalogs for benzonitrile in public databases did not contain hyperfine-splitting frequencies, and existing laboratory work at high resolution was limited to measurements below 11 GHz \cite{Wohlfart:2008hg}.  For these reasons, additional transitions of benzonitrile were measured in the laboratory at high resolution between 7 and 29 GHz to ensure the astronomical data could be interpreted \cite{Science:Materials}.  

\begin{figure}
\centering
\includegraphics[width=\textwidth]{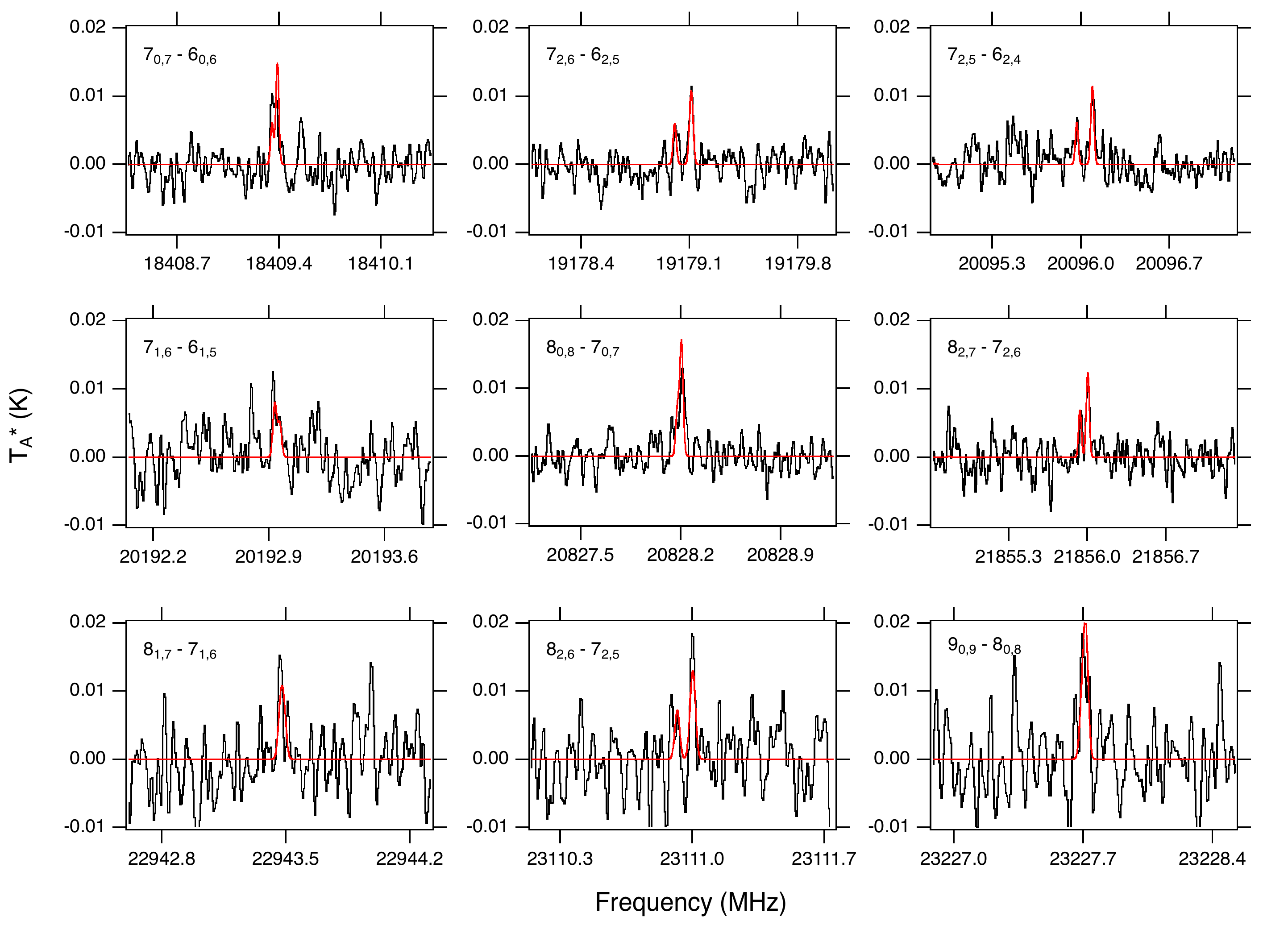}
\caption{\textbf{Detected emission lines of benzonitrile in TMC-1.}  Observational spectra are shown in black smoothed to a resolution of 5.7~kHz (0.08~km~s$^{-1}$) and shifted to a $v_{\rm{lsr}}$~=~5.83~km~s$^{-1}$.  A simulated spectrum of benzonitrile,  0.4~km~s$^{-1}$ linewidth, $N_T$~=~$4$~$\times$~$10^{11}$~cm$^{-2}$, and $T_{\rm{ex}}$~=~7~K is overlaid in red \cite{Science:Materials}.  Rotational quantum numbers are displayed in the upper left of each panel. The four transitions with well-resolved hyperfine structure are shown on an expanded frequency axis in \cite{Science:Materials}.}
\label{detections}
\end{figure}

Molecules in TMC-1 are typically well-described by a single excitation temperature between 5 and 10~K \cite{Gratier:2016fj,Loomis:2016jsa}.  Under these conditions, the strongest benzonitrile transitions fall between 20 and 40 GHz.  A total of 1.875 GHz of bandwidth was covered to high sensitivity ($T_A^*$~=~2~--~5~mK) between 18~--~23~GHz. In this range, eight of the nine strongest predicted rotational transitions were observed, each with a SNR~$\geq3$ (Fig.~\ref{detections}).  For six of these, characteristic $^{14}$N nuclear hyperfine splitting is partially or fully resolved for one or more components (Table~\ref{transitions}).  The emission features are best described by a $v_{\rm{lsr}}$~=~5.83~km~s$^{-1}$, a typical velocity for molecules in this source \cite{Kaifu:2004tk}.  All other strong transitions between 18 and 23~GHz fell into gaps in the spectral coverage, or in regions where insufficient noise levels were achieved. Taken together, these findings establish the presence of benzonitrile in TMC-1.

A joint analysis of all the lines yields a total column density $N_T$~=~$4$~$\times$~$10^{11}$~cm$^{-2}$ \cite{Science:Materials}, about twenty times less that of HC$_7$N (1.1~$\times$~$10^{13}$~cm$^{-2}$; \cite{Loomis:2016jsa}), an unsaturated linear cyanopolyyne with the same carbon and nitrogen composition as benzonitrile, in the same source.  Because the upper state energies of the observed transitions span only a narrow energy range (3.6 to 5.7~K), the excitation temperature could not be constrained from these observations.  Our analysis therefore assumed $T_{\rm{ex}}$~=~7~K, in the middle of the range of 5--10~K, derived from other molecules in this source \cite{Gratier:2016fj,Loomis:2016jsa}.   We also constrained the linewidth to 0.4~km~s$^{-1}$, based on the three fully-resolved hyperfine components. Although these components are some of the lowest SNR features, and our data are limited by the resolution of the observations (0.08~km~s$^{-1}$), this linewidth is consistent with that seen previously for other molecules in this source \cite{Kaifu:2004tk}.   Simulated spectra under these conditions are shown in Fig.~\ref{detections}, and are in agreement with the observations.

\begin{table}
\centering
\footnotesize
\caption{\textbf{Detected benzonitrile transitions.} Quantum numbers, frequencies, upper state energies ($E_U$), line strengths ($S_{ij}\mu^2$), observed intensities ($\Delta T_A^{\hspace{0.2em}*}$), and signal-to-noise ratio of detected benzonitrile transitions.  Statistical uncertainties (1$\sigma$), derived from the best-fitting constants in \cite{Science:Materials} are given in parentheses in units of the last significant digit.}
\begin{tabular}{c c c c c c c}
\hline\hline
\multicolumn{2}{c}{Transition}													&	Frequency		&	$E_U$		&	$S_{ij}\mu^2$		&	$\Delta T_A^{\hspace{0.2em}*}$$^{\dagger}$	& 					\\
$J^{\prime}_{K_a,K_c} - J^{\prime\prime}_{K_a,K_c}$	& $F^{\prime}-F^{\prime\prime}$	&	(MHz)			&	(K)			&	(Debye$^2$)		&	(mK)				\vspace{0.2em}			&	Signal-to-Noise		\\
\hline
$7_{0,7} - 6_{0,6}$								&	$6-5$					&	18409.3490(2)		&	3.61			&	39.5			&	\multirow{3}{*}{10.3(8)$^{b}$}				&	\multirow{3}{*}{4.4}	\\
											&	$8-7$					&	18409.3840(2)		&	3.61			&	52.9			&										&					\\
											&	$7-6$					&	18409.3879(2)		&	3.61			&	45.7			&										&		\vspace{1em}\\

$7_{2,6} - 6_{2,5}$\vspace{0.5em}					&	$7-6$					&	19179.0017(2)		&	4.52			&	42.4			&	5.4(5)								&	2.7				\\
											&	$6-5$					&	19179.1027(2)		&	4.52			&	36.6			&	\multirow{2}{*}{10.6(5)$^{b}$}				&	\multirow{2}{*}{5.4}	\\
											&	$8-7$					&	19179.1128(2)		&	4.52			&	49.1			&										&		\vspace{1em}\\
											
$7_{2,5} - 6_{2,4}$\vspace{0.5em}					&	$7-6$					&	20095.9645(2)		&	4.62			&	42.5			&	6.6(5)								&	3.3				\\
											&	$6-5$					&	20096.0824(2)		&	4.62			&	36.7			&	\multirow{2}{*}{10.0(2)$^{b}$}				&	\multirow{2}{*}{4.9}	\\
											&	$8-7$					&	20096.0917(2)		&	4.62			&	49.2			&										&		\vspace{1em}\\
											
$7_{1,6} - 6_{1,5}$								&	$6-5$					&	20192.9325(2)		&	4.11			&	39.0			&	\multirow{3}{*}{8(1)$^{b}$}					&	\multirow{3}{*}{2.1}	\\
											&	$7-6$					&	20192.9342(2)		&	4.11			&	45.1			&										&					\\
											&	$8-7$					&	20192.9632(2)		&	4.11			&	52.2			&										&		\vspace{1em}\\
											
$8_{0,8} - 7_{0,7}$								&	$7-6$					&	20828.1746(2)		&	3.20			&	46.2			&	\multirow{3}{*}{11.7(9)$^{b}$}				&	\multirow{3}{*}{5.9}	\\
											&	$9-8$					&	20828.2012(2)		&	3.20			&	59.6			&										&					\\
											&	$8-7$					&	20828.2045(2)		&	3.20			&	52.5			&										&		\vspace{1em}\\		
											
$8_{2,7} - 7_{2,6}$\vspace{0.5em}					&	$8-7$					&	21855.9322(3)		&	5.57			&	49.7			&	7.1(4)								&	3.1				\\
											&	$7-6$					&	21855.9944(3)		&	5.57			&	43.8			&	\multirow{2}{*}{10.7(5)$^{b}$}				&	\multirow{2}{*}{4.7}	\\
											&	$9-8$					&	21856.0064(3)		&	5.57			&	56.4			&										&		\vspace{1em}\\
											
$8_{1,7} - 7_{1,6}$								&	$7-6$					&	22943.4640(3)		&	5.21			&	45.8			&	\multirow{3}{*}{16.2(8)$^{b}$}				&	\multirow{3}{*}{3.2}	\\
											&	$8-7$					&	22943.4729(3)		&	5.21			&	52.0			&										&					\\
											&	$9-8$					&	22943.4885(3)		&	5.21			&	59.1			&										&		\vspace{1em}\\
											
$8_{2,6} - 7_{2,5}$\vspace{0.5em}					&	$8-7$					&	23110.9171(3)		&	5.73			&	49.9			&	6.2(9)								&	1.2				\\
											&	$7-6$					&	23110.9923(3)		&	5.73			&	43.9			&	\multirow{2}{*}{19.6(9)$^{b}$}				&	\multirow{2}{*}{3.8}	\\
											&	$9-8$					&	23111.0042(3)		&	5.73			&	56.7			&										&		\vspace{1em}\\
											
$9_{0,9} - 8_{0,8}$								&	$8-7$					&	23227.6903(3)		&	5.72			&	52.9			&	\multirow{3}{*}{16(2)$^{b}$}				&	\multirow{3}{*}{3.3}	\\
											&	$10-9$					&	23227.7105(3)		&	5.72			&	66.3			&										&					\\
											&	$9-8$					&	23227.7127(3)		&	5.72			&	59.3			&										&					\\
\hline
\multicolumn{7}{l}{$^{\dagger}$Uncertainty in the Gaussian fit. A conservative 30\% uncertainty in the absolute flux calibrated}\\
\multicolumn{7}{l}{  value is assumed \cite{Science:Materials}.}\\
\multicolumn{7}{l}{$^{b}$Indicates blended hyperfine components; $\Delta T_A^{\hspace{0.2em}*}$ is the peak value of the observed feature.}\\
\end{tabular}
\label{transitions}
\end{table}

\begin{figure}[h!]
\centering
\includegraphics[width=0.65\textwidth]{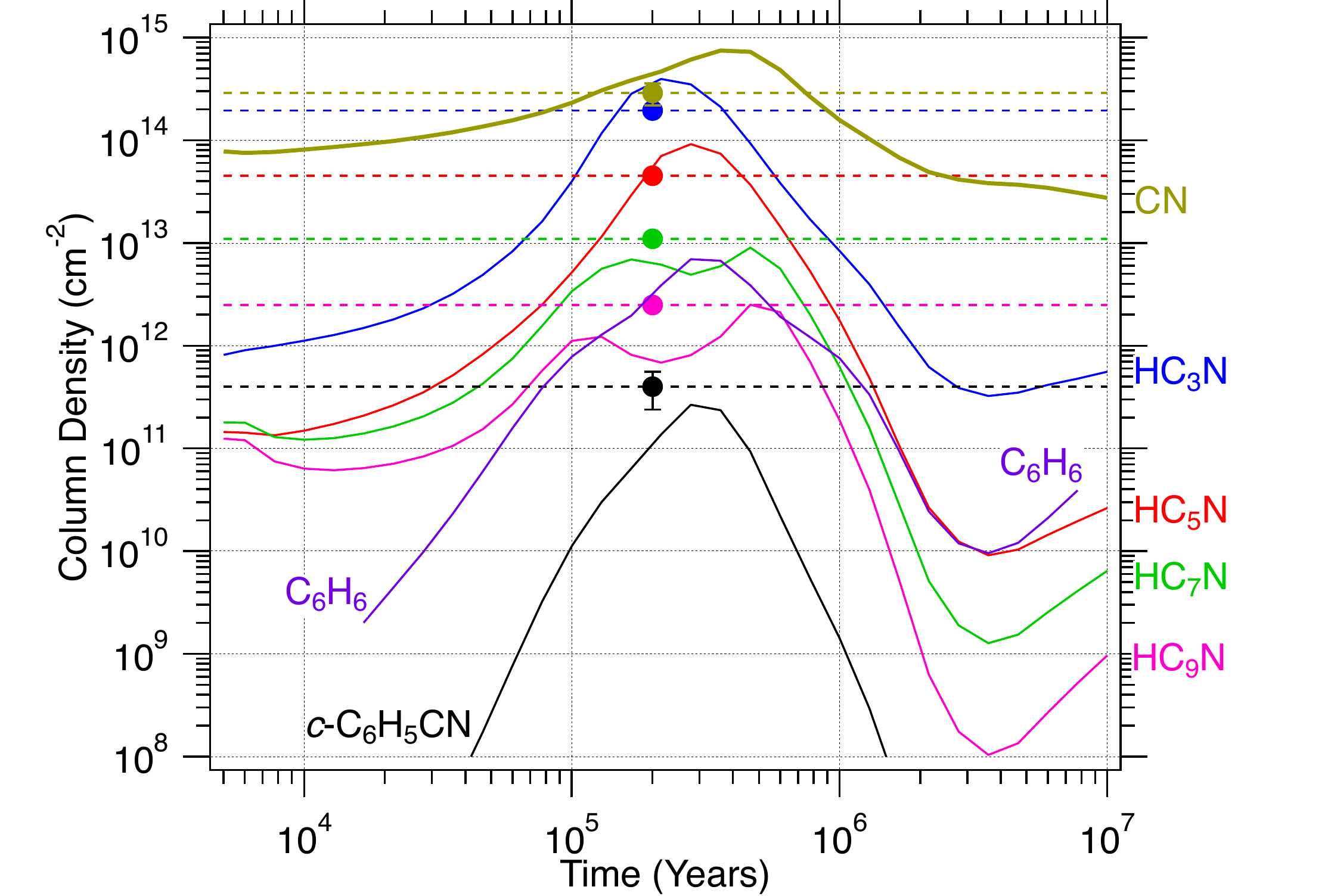}
\caption{\textbf{Chemical model of TMC-1.} Results of a three-phase astrochemical model, updated to include the benzonitrile formation pathway given in Reaction~\ref{bn1}.  The model gas-phase column densities as a function of time are given as solid, colored lines.  Column densities of benzonitrile, cyanopolyynes, and CN derived from observations are shown as  dots with dashed horizontal lines (from top to bottom: CN, HC$_3$N, HC$_5$N, HC$_7$N, HC$_9$N, and benzonitrile \cite{Loomis:2016jsa,Crutcher:1984jo}).  The derived column densities are plotted at the chemical age of TMC-1 ($\sim2\times10^5$ yrs; \cite{hincelin2011}).}
\label{model}
\end{figure}

The pathways leading to the formation of benzonitrile at low temperature and in low density environments have not been studied in detail. Perhaps the only promising astrochemically-relevant formation pathway discussed in the literature is the neutral-neutral reaction 
\begin{equation}
   \rm{CN} + \rm{\textnormal{\emph{c}-}C}_6\rm{H}_6 \rightarrow \rm{\textnormal{\emph{c}-}C}_6\rm{H}_5\rm{CN} + \rm{H.}
    \label{bn1}
\end{equation}	
This barrierless, exothermic reaction has been considered previously \cite{woon2006,Trevitt:2010ep}. In an effort to determine the contribution of Reaction~\ref{bn1} to the observed abundance of benzonitrile in TMC-1, we have modified the Kinetic Database for Astrochemistry (KIDA) gas-phase reaction network \cite{Wakelam:2015dr} to include this reaction, as well as destruction pathways from photons, ions, and depletion onto grains \cite{Science:Materials}. This network was then combined with the \texttt{NAUTILUS-1.1} modeling code \cite{Ruaud:2016bv} assuming elemental abundances and physical conditions appropriate for TMC-1 \cite{hincelin2011}.  A number of additional gas-phase formation routes for the precursor benzene were also considered, and included in the modified network \cite{Science:Materials}.  A column density of H$_2$~=~10$^{22}$~cm$^{-2}$ \cite{Gratier:2016fj} was used to convert from modeled abundances to column densities to compare with observations.  Figure~\ref{model} shows the derived column densities and those predicted by the model for benzonitrile, benzene, CN, and the cyanopolyynes HC$_3$N, HC$_5$N, HC$_7$N, and HC$_9$N. While the calculated column densities of most of the cyanopolyynes agree with observational results within a factor of two, the predicted benzonitrile column density is smaller than the derived value by nearly a factor of four.  

That difference may be explained by other formation routes for benzene, benzonitrile, or both that are not considered in our model. For instance, experiments have found that benzene can be formed in electron-irradiated acetylene ices \cite{zhou2010}. In astrochemical models, the addition of cosmic ray driven irradiation chemistry in the solid phase has been found to improve agreement between observational and theoretical abundances for other large interstellar molecules \cite{abplanalp2016}, although for such grain-surface processes to contribute to gas-phase abundances there must exist efficient non-thermal desorption mechanisms. Recent theoretical and experimental work suggests that interactions between cosmic rays and grain surfaces could result in the liberation of solid-phase species into the gas phase via processes that are viable in cold cores such as TMC-1 \cite{ribeiro2015}.

Benzene is also known to be produced in irradiated acetylene gas \cite{field1964}. Radiation chemistry differs from photochemistry in a number of ways \cite{wishart1998} and may be a viable formation pathway for aromatic and PAH molecules as a result of their increased photostability (relative to simpler organics \cite{Jochims:1999df}).  However, there is insufficient theoretical and experimental work to include such pathways in our model.  Previous work also suggests a possible connection between benzonitrile and the cyanopolyynes \cite{Loomis:2016jsa} . That work showed a sharp decrease in the calculated abundance of HC$_{11}$N relative to the abundance trend of the $n (\rm{odd}) =3-9$ cyanopolyynes, possibly due to cyclization processes for HC$_n$N molecules which could eventually lead to functionalized aromatics such as benzonitrile. 

We also consider the possibility that benzonitrile itself may be a contributor to the UIR bands.  The vibrational spectrum of benzonitrile has been studied in the infrared, both experimentally \cite{Green:1976dd} and theoretically \cite{Csaszar:1989fq}, but it does not appear in spectral databases \cite{Boersma:2014bb}, and it is not commonly considered as a potential UIR carrier.  Nevertheless, the interstellar IR emission features at 3.3 $\mu$m (C--H aromatic stretch) and 4.48 $\mu$m (C$\equiv$N stretch) are both in agreement with very strong IR modes of benzonitrile \cite{Csaszar:1989fq}, thus making it a potential carrier in its own right, as well as a likely precursor to polyaromatic species.

In summary, we have detected the aromatic molecule benzonitrile in TMC-1, using radio astronomy to probe this class of molecules.  These species may be either direct contributors to the UIR bands, or precursors to the carriers themselves.

\clearpage




\begin{scilastnote}
\item[] \textbf{Supplementary Materials}

\vspace{-1em}
www.sciencemag.org

\vspace{-1em}
Materials and Methods

\vspace{-1em}
Figures S1, S2

\vspace{-1em}
Tables S1, S2, S3, S4, S5

\vspace{-1em}
References (40-58)

\item[] \textbf{Acknowledgements}
\item[] Support for B.A.M. was provided by NASA through Hubble Fellowship grant \#HST-HF2-51396 awarded by the Space Telescope Science Institute, which is operated by the Association of Universities for Research in Astronomy, Inc., for NASA, under contract NAS5-26555.  A.M.B. is a Grote Reber Fellow, and support for this work was provided by the NSF through the Grote Reber Fellowship Program administered by Associated Universities, Inc./National Radio Astronomy Observatory. S.V.K. acknowledges support from Basic Research Program P-7 of the Presidium of the Russian Academy of Sciences. E.H. and C.N.S. acknowledge support from the National Science Foundation. M.C.M. acknowledges support from NSF grant AST-1615847.  The authors thank G.A. Blake for a critical reading of the manuscript and J. Mangum for helpful discussions. The National Radio Astronomy Observatory is a facility of the National Science Foundation operated under cooperative agreement by Associated Universities, Inc.  The Green Bank Observatory is a facility of the National Science Foundation operated under cooperative agreement by Associated Universities, Inc. \\

The Nobeyama observations are archived at \url{www.cv.nrao.edu/PRIMOS}, and the GBT observations at \url{https://archive.nrao.edu/archive/advquery.jsp} under IDs AGBT02C\_012, AGBT17A\_164, and AGBT17A\_434. Laboratory data are tabulated in the Supplementary Material. The modifications we applied to KIDA and the output of our chemical model are at \url{doi:10.18130/V3/4IFDBP}.

\clearpage

Supplementary Materials for

\title{Detection of the Aromatic Molecule Benzonitrile ($c$-C$_6$H$_5$CN) in the Interstellar Medium}

\author
{Brett A. McGuire,$^{1,2,\ast}$ Andrew M. Burkhardt,$^{3}$ Sergei Kalenskii,$^4$\\ Christopher N. Shingledecker,$^5$ Anthony J. Remijan,$^1$  Eric Herbst,$^{5,3}$\\ and Michael C. McCarthy$^{2,6}$\\
\\
\normalsize{$^{1}$National Radio Astronomy Observatory, Charlottesville, VA 22903, USA}\\
\normalsize{$^{2}$Harvard-Smithsonian Center for Astrophysics, Cambridge, MA 02138, USA}\\
\normalsize{$^{3}$Department of Astronomy, University of Virginia, Charlottesville, VA 22904 USA}\\
\normalsize{$^{4}$Astro Space Center, Lebedev Physical Institute, Russian Academy of Sciences, Moscow, Russia}\\
\normalsize{$^{5}$Department of Chemistry, University of Virginia, Charlottesville, VA 22904 USA}\\
\normalsize{$^{6}$School of Engineering and Applied Sciences, Harvard University, Cambridge, Massachusetts 02138 USA}\\
\\
\normalsize{$^\ast$To whom correspondence should be addressed; E-mail: bmcguire@nrao.edu.}
}

\item[] \textbf{This PDF file includes:}


\vspace{-1em}
Materials and Methods

\vspace{-1em}
Figures S1, S2

\vspace{-1em}
Tables S1, S2, S3, S4, S5

\vspace{-1em}
References (40-58)

 \end{scilastnote}
 
 \clearpage
 
 \part*{Materials \& Methods}

\renewcommand{\thefigure}{S\arabic{figure}}
\renewcommand{\thetable}{S\arabic{table}}
\renewcommand{\theequation}{S\arabic{equation}}
\setcounter{figure}{0}
\setcounter{table}{0}
\setcounter{equation}{0}

\section*{Observations}

A detailed description of the GBT observations is given in \cite{McGuire:2017ud}.  The observations of TMC-1 using the 100 m Robert C. Byrd Green Bank Telescope in Green Bank, West Virginia were centered on right ascension =~04$^h$41$^m$42.5$^s$, declination =~25$^{\circ}$41$^{\prime}$27.0$^{\prime\prime}$ (J2000 equinox).  Pointing observations were conducted every hour; the pointing accuracy is estimated to be within 2$^{\prime\prime}$.  The beam size varied from 32--40$^{\prime\prime}$ across the observed frequency range.  The K-band Focal Plane Array \cite{kband} was used with the VEGAS spectrometer backend \cite{Roshi:2012he} configured to provide 187.5 MHz total bandwidth in each of ten target windows at a 1.4 kHz (0.02 km s$^{-1}$) spectral resolution. Two observing setups were used, as the VEGAS backend can support at most eight simultaneous frequency windows.  Observations were conducted in position-switching mode, using a 1$^{\circ}$ offset throw, with 120 s of integration at each position and between $\sim$7.5 and 15~hours of total on-source integration, depending on the frequency window.  The resulting spectra were placed on the atmosphere-corrected antenna temperature ($T_A$*) scale \cite{Ulich:1976yt}.  Data reduction was performed using the GBTIDL software package \cite{gbtidl}, and archival data of TMC-1 from Project GBT02C\_012 (PI: J. Hollis) were included to increase the SNR. The spectra were averaged using a weighting scheme which corrects for the measured system temperature ($T_{sys}$) during each 240 s on-off position cycle.  The spectra were smoothed to a resolution of 5.7 kHz (0.08~km~s$^{-1}$), sufficient to provide $\geq$4 points across each 0.4~km~s$^{-1}$ full-width half-maximum (FWHM).  A polynomial fit was used to correct for baseline fluctuations.  The final root-mean-square (RMS) noise varied from 2 to 5 mK across the observations.

The Nobeyama 45-m observations from 8.8~--~50~GHz were conducted over a period of 13 years from 1984 -- 1997, and are described in detail in \cite{Kaifu:2004tk}.  The RMS noise varied across the survey, but was typically $\sim$10~mK in 20~kHz channels.

\section*{Column Density Calculations}

The overall data analysis procedure follows the general methodology used in previous observations \cite{McGuire:2016ba,McGuire:2017ud}.  The column density was determined using the formalism of \cite{Hollis:2004uh}:
\begin{equation}
N_T = \frac{Qe^{E_u/T_{ex}}}{\frac{8\pi^3}{3k}\nu S_{ij}\mu^2g_I}\times \frac{\frac{1}{2}\sqrt{\frac{\pi}{\ln(2)}}\frac{\Delta T_A\Delta V}{\eta_B}}{1 - \frac{e^{h\nu/kT_{ex}} -1}{e^{h\nu/kT_{bg}} -1}}
\label{column}
\end{equation}
where $N_T$ is the column density (cm$^{-2}$), $E_u$ is the upper state energy (K), $\Delta T_A \Delta V$  is integrated line intensity (K~cm~s$^{-1}$), $T_{ex}$ is the excitation temperature (K), $T_{bg}$ is the background continuum temperature (2.7 K), $\nu$ the transition frequency (Hz), $S_{ij}$ is the intrinsic line strength, $\mu^2$ is the transition dipole moment (Debye), and $\eta_B$ is the beam efficiency ($\sim$0.92 for the GBT at 20 GHz). An additional factor of $g_I$ has been added to correct for the spin statistical weights (see below).  The partition function, $Q$, is discussed in detail below. We assume that the source fills the beam (see \cite{Loomis:2016jsa}).  The column density was calculated using the highest SNR transition that was at least partially hyperfine-resolved, the $F$~=~6--5 and 8--7 components of the $7_{2,5} - 6_{2,4}$ transition; the value matches the observed spectra quite well by visual inspection, and certainly within the uncertainties. An alternative formulation of Equation~\ref{column}, along with a detailed description of the derivations, is given in Mangum \& Shirley \cite{Mangum:2015wp}.  

\subsection*{Partition Functions}

The total partition function $Q$ is given by $Q = Q_{\rm{vib}}\times Q_{\rm{rot}}$. The rotational partition function, $Q_{\rm{rot}}$, is calculated explicitly by direct summation of states using Equation \ref{qrot} (c.f. \cite{Gordy:1984uy}; $Q_{\rm{rot}}$[7~K]~=~480).  Here, $\sigma = 1$ for an asymmetric molecule.
\begin{equation}
Q_r = \frac{1}{\sigma}\sum_{J = 0}^{J = \infty}\sum_{K = -J}^{K = J}(2J+1)g_Ie^{-E_{J,K}/kT_{ex}}
\label{qrot}
\end{equation}
The value of $g_I$ is found following the definition \cite{Gordy:1984uy} as given in Equation~\ref{gi}.
\begin{equation}
g_I \equiv \frac{g_{nuclear}}{g_n}
\label{gi}
\end{equation}
Here, $g_{nuclear}$ arises from the combining the two sets of equivalent nuclei which result in an overall symmetry or asymmetry.  The symmetric and asymmetric ($K = \rm{even}$ and $K = \rm{odd}$) values and the nuclear spin weights are calculated in the final line for benzonitrile ($I_1 = I_2 = \frac{1}{2}$) using \cite{Gordy:1984uy}.

\begin{eqnarray}
g^s_{nuclear} 	&=& \psi_{1,sym}\psi_{2,sym} + \psi_{1,asym}\psi_{2,asym} \nonumber \\
			&=& (I_1 + 1)(2I_1 +1)(I_2 +1)(2I_2 + 1) + I_1(2I_1 +1)I_2(2I_2 +1) \nonumber \\
			&=& 10 
\label{gsym}
\end{eqnarray}
\begin{eqnarray}
g^a_{nuclear} 	&=& \psi_{1,asym}\psi_{2,sym} + \psi_{1,sym}\psi_{2,asym} \nonumber \\
			&=& (I_1+1)(2I_1+1)I_2(2I_2+1) + I_1(2I_1+1)(I_2+1)(2I_2+1) \nonumber \\
			&=& 6 
\label{gasym}
\end{eqnarray}
To solve for $g_I$, $g_{nuclear}$ must be divided by $g_n$ given by Equation~\ref{gn}, which for benzonitrile is 16.  Thus, the final values for $g_I$ requires division by $(2I +1)^4$.  That makes $g_{I,even} = 5/8$ and $g_{I,odd} = 3/8$.
\begin{equation}
g_n = \prod_i\prod_n(2I_i+1)
\label{gn}
\end{equation}

We have calculated the energies of the vibrational states of benzonitrile at the WB97XD/6-311++G(d,p) level of theory and basis set to determine their possible contribution to the overall partition function at these temperatures.  The vibrational contribution is given by Equation~\ref{qvib}.
\begin{equation}
Q(T)_{vib}=\prod\limits_{\substack{i=1}}^{3N-6} \frac{1}{1-e^{-E_i/kT}}
\label{qvib}
\end{equation}
The lowest energy levels are at 162~cm$^{-1}$ (233~K) and 184~cm$^{-1}$ (265~K).  At these temperatures, the vibrational contribution to the partition function is less than 10$^{-14}$.  Indeed, the vibrational partition function does not contribute at the 1\% level until $T_{\rm{ex}} \sim$38~K.

\subsection*{Uncertainties}

We estimate an overall uncertainty of 40\% in the derived column density for benzonitrile.  This is based on a number of contributing factors, enumerated below, and added in quadrature:

\begin{enumerate}
\item A 30\% uncertainty arising from the absolute flux uncertainty in the observations is assumed.  This incorporates the 20\% uncertainty in the strongest transitions arising from the noise of the observations (SNR$\sim$5), and a (conservative) 20\% calibration uncertainty, added in quadrature.
\item A 5--10\% uncertainty from the Gaussian fits to the line profiles.
\item A 20\% contribution arising from the accuracy of the linewidth used (0.40~km~s$^{-1}$) compared to the resolution element of the observations ($\sim$0.08~km~s$^{-1}$).
\item A 20\% contribution arising from the choice of excitation temperature (7~K) compared to the lower and higher ends of the range commonly seen in the source (5--10~K).
\end{enumerate}

We further note that the uncertainties given in Table \ref{transitions} are the purely-statistical 1$\sigma$ standard deviation uncertainty in the Gaussian fitting routine used to determine the peak value of $T_A^*$ ($\sim$5--10\% as noted above).  The SNR is then calculated by simply dividing this peak value by the RMS of the data in that region.  The actual uncertainty in the peak is significantly higher, due to the other contributions noted above.

\subsection*{Residuals and Goodness of Fit}

\begin{figure}
\centering
\includegraphics[width=\textwidth]{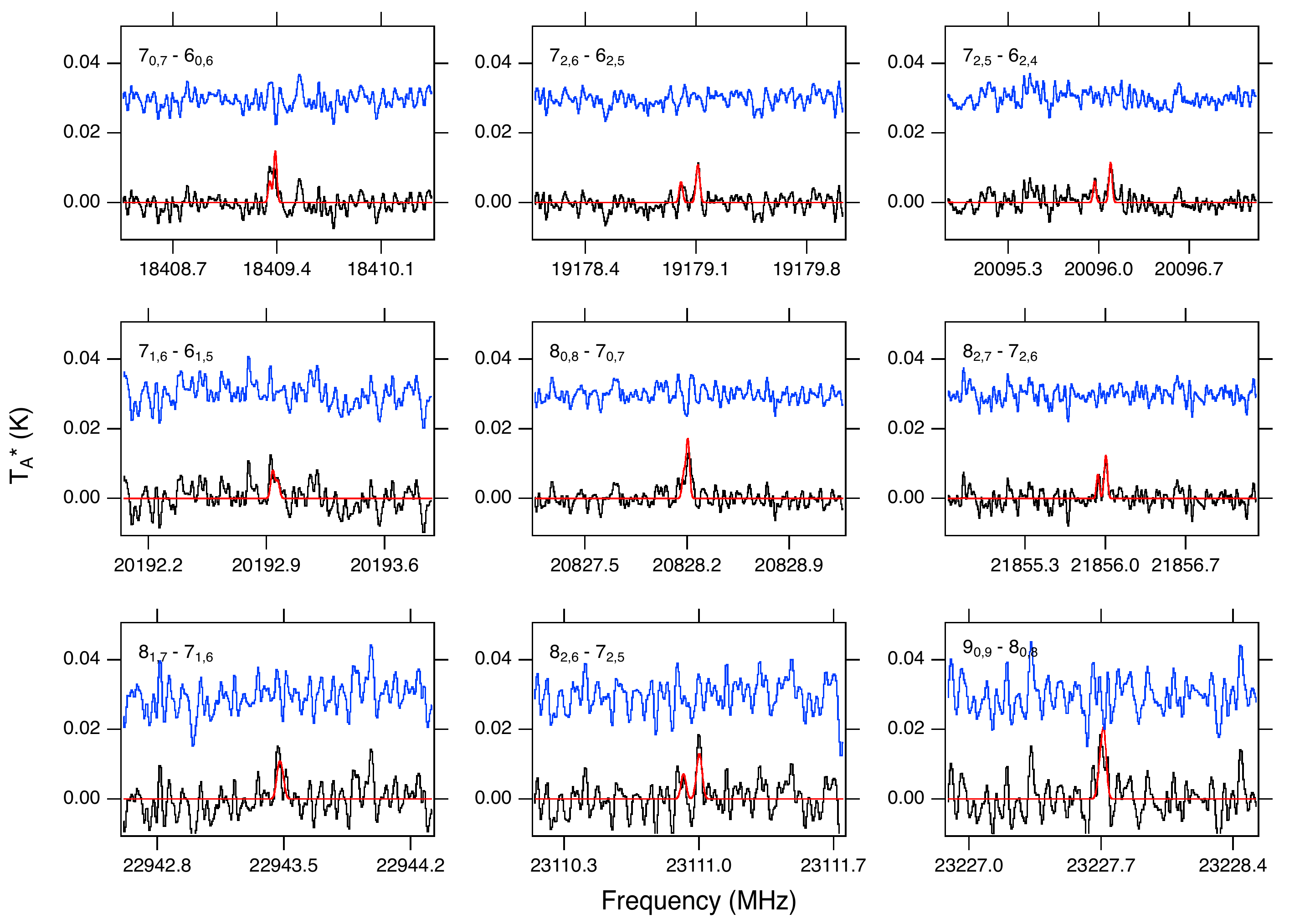}
\caption{\textbf{Fit residuals.} Residuals for the nine detected transitions are shown in the upper blue tracer, and are consistent with the baseline noise level of the observations in each window.  All other parameters are the same as in Figure~\ref{detections}.}
\label{residuals}
\end{figure}

\begin{figure}
\centering
\includegraphics[width=0.85\textwidth]{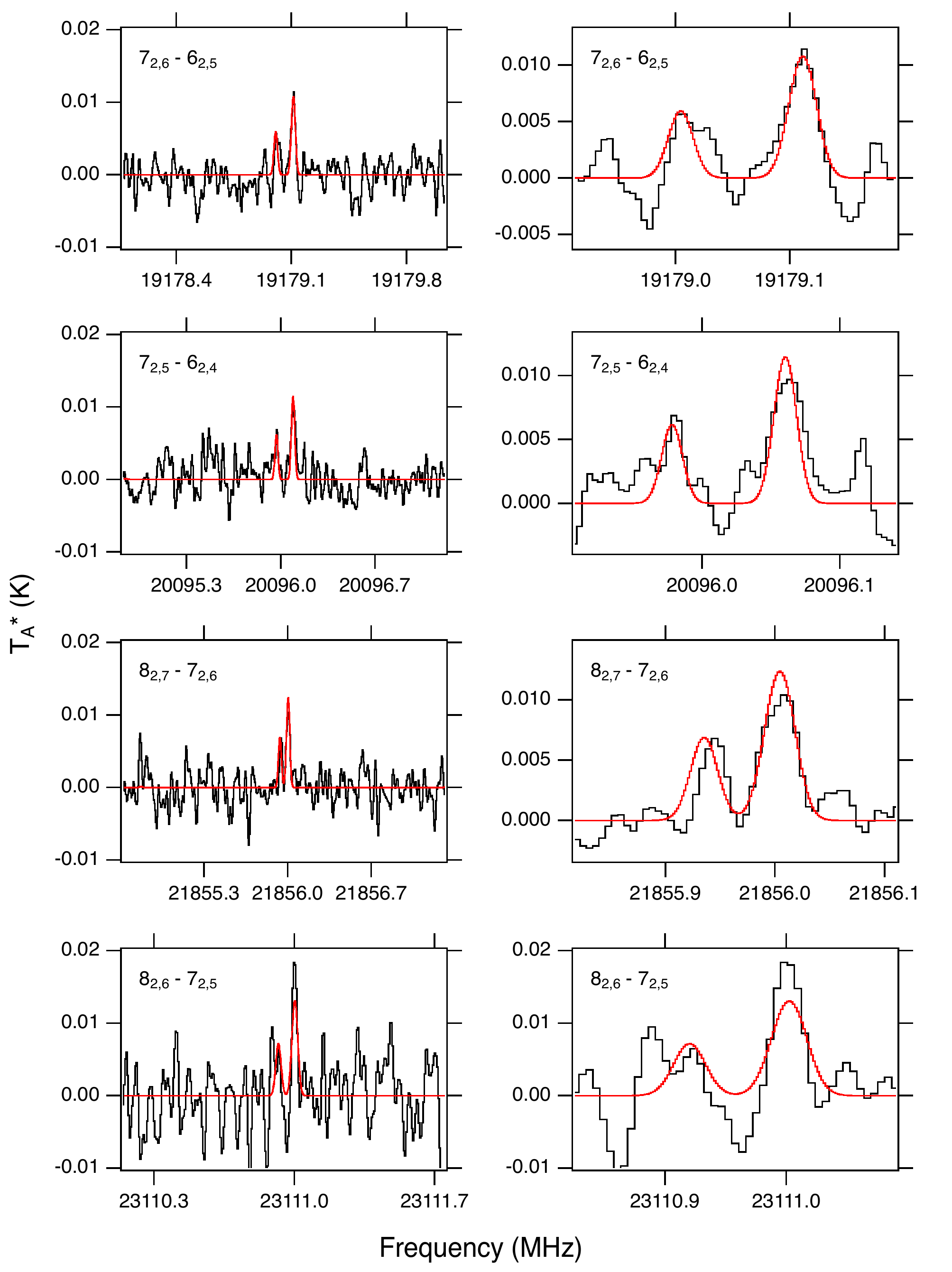}
\caption{\textbf{Detailed spectral comparisons.} Comparison of the model spectra (red) to the observed spectra (black) on a zoomed-in scale to show detail on the four detected transitions with a well-resolved hyperfine component.}
\label{zoomed}
\end{figure}

The residuals for the nine detected transitions are shown in Figure~\ref{residuals}, and are consistent with the baseline noise level of the observations in each window.  Figure~\ref{zoomed} shows the comparison of the model spectra to the observed spectra on a zoomed-in scale to show greater detail for the four transitions with a well-resolved hyperfine component.  The $8_{2,7} - 7_{2,6}$ $F = 8-7$ component at 21855.93 MHz is the poorest agreement in line-center position of any detected transition.  The difference is 7.2(4) kHz, equivalent to 1.25(5) channel widths.  Given the SNR of the line, and the agreement of every other transition, this is likely a simple issue of the noise slightly affecting the line center position in the Gaussian fit.

\subsection*{Composite Averages}

			The process of building a composite average (CA) used in this work consists of the following steps:
			
				\begin{enumerate}
				
					\item Transition frequencies for target molecules are determined, and the broadband observational spectra are trimmed into narrow-band elementary spectra centered on those frequencies.
					
					\item A single excitation temperature is assumed (7~K in the case of TMC-1) and the strongest predicted line with observational coverage is set as the reference line.
					
					\item Each elementary spectrum is multiplied by the ratio of the brightness temperatures of the given transition relative to that of the reference line.  As a result, the brightness temperatures of all lines are normalized, while the RMS noise of the elementary spectra (except for the reference spectrum) increases.
					
					\item  Finally, the elementary spectra are averaged together using weights inversely proportional to their RMS noise levels, and the final CA is obtained.
				
				\end{enumerate}

\section*{Benzene Formation Chemistry}

A number of formation pathways have been proposed for the key precursor species, benzene, in interstellar regions. One of the first is the dissociative recombination reaction \cite{mcewan1999}:
\begin{equation}
   \rm{\textnormal{\emph{c}-}}\mathrm{C_6H_7^+} + e^- \rightarrow \rm{\textnormal{\emph{c}-}}\mathrm{C_6H_6} + \mathrm{H}
   \label{benzene1}
\end{equation}
which is the main formation route for benzene in the KIDA network. Another production pathway that could occur under interstellar conditions is the neutral-neutral reaction between the ethynyl radical and 1,3-butadiene \cite{jones2011}:

\begin{equation}
    \mathrm{C_2H} + \mathrm{H_2CCHCHCH_2} \rightarrow \rm{\textnormal{\emph{c}-}}\mathrm{C_6H_6} + \mathrm{H}.
    \label{benzene2}
\end{equation}
This reaction satisfies the typical requirement for gas-phase chemistry in cold cores such as TMC-1 since it is thought to be barrierless, exothermic, and can occur on a single-collision basis. Given those characteristics which are ideal for interstellar gas-phase chemistry, we have added this reaction, with a rate coefficient of $\sim$$3\times10^{-10}$ cm$^3$ s$^{-1}$ (c.f. \cite{jones2011}),  to our modified version of the KIDA 2014 reaction network \cite{Wakelam:2015dr} to which we have previously added reactions of the HC$_n$O ($n=3-7$) family of molecules \cite{McGuire:2017ud}.

There is another formation route for benzene that has been previously studied in the context of combustion chemistry \cite{miller2003,georgievskii2007} involving the reaction between two propargyl radicals:
\begin{equation}
    \mathrm{C_3H_3} + \mathrm{C_3H_3} \rightarrow \rm{\textnormal{\emph{c}-}}\mathrm{C_6H_6^*}
    \rightarrow \rm{\textnormal{\emph{c}-}}\mathrm{C_6H_5} + \mathrm{H}.
    \label{benzene3}
\end{equation}
Here, the benzene is produced in an unstable excited state and will dissociate into atomic hydrogen and the phenyl radical unless it is stabilized radiatively or via collision with a third-body. However, three-body collisions in molecular clouds occur on timescales greater than the lifetime of the cloud \cite{kaiser2002}. Moreover the rate coefficient for the radiative stabilization of the C$_6$H$_6$ product is estimated to be several orders of magnitude slower than its dissociation rate coefficient \cite{jones2011}. Thus, reaction \eqref{benzene3} is unlikely to contribute to the abundance of benzene in a cold core such as TMC-1. 

\section*{Laboratory Measurements of Benzonitrile, Spectroscopic Constants, and Comparisons to Previous Work} 

\medskip

Benzonitrile is a closed-shell asymmetric top moderately close the prolate limit [Ray's Asymmetry Parameter: $\kappa=(2B-A-C)/(A-C)=-0.85027$].  Owing to its $C_{2v}$ symmetry and two sets of equivalent fermions, it only has a non-zero dipole moment along its principal ($a$) inertial axis, and transitions with $K_a$ even and odd have statistical weights of 5 and 3, respectively. Its rotational spectrum has been reported at high spectral resolution by several research groups~\cite{fliege:1124,vormann:283,Wohlfart:2008hg,Wlodarczak:1989vk}.  The most recent and accurate measurements observed six hyperfine-split rotational transitions up to $J=4$ and $K_a=1$ between 2.8 and 10.9~GHz using Fourier-Transform (FT) microwave spectroscopy \cite{Wohlfart:2008hg}; estimated experimental uncertainties for the individual hyperfine lines are quite small, ranging from 0.5 to 2\,kHz.  By combining this data with higher-frequency measurements from Refs.~\cite{fliege:1124} and~\cite{vormann:283}, which were assigned uncertainties of 20 and 5\,kHz, respectively, the three rotational constants and all five quartic centrifugal distortion constants were determined. A precise dipole moment for benzonitrile [4.5152(68)\,D] has been derived by performing Stark measurements on three rotational transitions \cite{Wohlfart:2008hg}.

\medskip

Because molecules in TMC-1 are characterized by $T_{ex}\sim5-10$\,K, the most intense lines of benzonitrile are predicted to lie in the 20--40 GHz range, and to arise from transitions between both higher $J$ and $K_a$ levels.  For this reason we have extended the rotational measurements of this molecule to higher frequency in the laboratory, with two goals in mind: i) to measure rest frequencies for all of the astronomically most intense lines predicted in TMC-1, and ii) to derive spectroscopic constants so that the rotational spectrum of this molecule can be predicted throughout the centimeter-wave band.  The precision required is better than 2~kHz, equivalent to a radial velocity of 0.06\,km~s$^{-1}$ (at~20 GHz), or a fraction of the channel width ($\sim$5\,kHz) typically used for observations of TMC-1 because  spectral features there are so narrow.

\medskip

Including hyperfine-splitting, nearly 150 lines of benzonitrile originating from more than 30 rotational transitions (Table~\ref{bn:lines}) have been measured between 7 and 29\,GHz by FT microwave spectroscopy in a Fabry-Perot cavity~\cite{mccarthy:105,mccarthy:611}. This data set includes transitions up to $J = 11$ and $K_a= 6$. Because benzonitrile has a substantial vapor pressure at standard temperature and pressure (191\,mm Hg), it was possible to entrain and  heavily dilute its vapor in a neon buffer gas, and then adiabatically expand this gas mixture using a pulsed value through a small hole in one of the two Fabry-Perot cavity mirrors.  As the gas expands into the vacuum chamber the rotational temperature drops  rapidly, to a few degrees K near the beam waist of the cavity. Linewidths are sharp in this arrangement, about 5\,kHz FWHM for closed-shell molecules like benzonitrile.  By applying a short microwave pulse as the gas passes through the beam waist, and then monitoring the resulting free induction decay (FID), it is possible to detect faint signals using a sensitive microwave receiver.  For high resolution measurements such as these, the FID was sampled every 500\,nsec for approximately 2000\,$\mu$s, equivalent to a spacing of 1\,kHz in the frequency domain.  Rest frequencies were determined by adding the small frequency offset ($<$1 MHz) of the Doppler doublet to the synthesizer (pump) frequency, which was locked to a highly stable ($< 5 \times 10^{-9}$) Rb frequency standard.  The frequency offset was derived using a least-squares fitting procedure which takes into account both Doppler components.  When comparison is possible, rest frequencies derived from this work and Ref.~\cite{Wohlfart:2008hg} agree to within 1\,kHz.

\medskip

Table~\ref{bn:constants} provides a comparison of the newly-derived spectroscopic constants, derived using the \textsc{spfit/spcat} suite of programs~\cite{pickett:371}, using a Watson-$A$ Hamiltonian in the I$^r$ representation, with those of Ref.~\cite{Wohlfart:2008hg}, and as well as those determined using a global fit which combines the two sets of high-resolution FT microwave measurements.  As indicated in Table~\ref{bn:compare1}, differences in the two sets of constants are not statistically significant: seven of the eight constants in Ref.~\cite{Wohlfart:2008hg} agree to within 0.7$\sigma$ (0.7 standard deviations) to the newly derived ones, and the eighth, a hyperfine-splitting constant $\chi_{aa}$(N), only differs by 1.7$\sigma$. Because the new data includes transitions between higher $J$ and $K_a$ levels, however, the $A$ rotational constant is roughly 10 times more precise, and the precision of the five quartic centrifugal distortion constants has been improved likewise, by factors of 2-8 compared to those reported in Ref.~\cite{Wohlfart:2008hg} (Table~\ref{bn:compare1}); the low fit RMS error (1.2\,kHz) and weighted average (0.62) both indicate that no additional hyperfine or centrifugal terms are required.  Despite the large number of additional transitions measured, benzonitrile appears to be a fairly rigid molecule, as its cm-wave rotational spectrum is still well described by a Hamiltonian which includes only quartic centrifugal distortion terms.  An improved set of spectroscopic constants for benzonitrile is obtained by including several low-$J$ transitions in Ref.~\cite{Wohlfart:2008hg} with our data set. Although doing so has little or no effect on the higher-frequency lines of astronomical interest, it does improve the precision of the two nitrogen quadrupole coupling constants by nearly a factor of two (see Table~\ref{bn:compare2}). 

\medskip

On the basis of the new laboratory measurements, hyperfine-split rotational transitions of benzonitrile can  be predicted to better than 2~kHz, or 0.1~km~s$^{-1}$ below 30~GHz.  The formal statistical uncertainty of rest frequencies predicted from the best-fitting constants in Table~\ref{measurements} is only a few 0.1~kHz, or about 0.02~km~s$^{-1}$ below 30~GHz.  This level of accuracy and precision is sufficient to predict closely-spaced hyperfine structure, and more than adequate for cold, quiescent sources such as TMC-1, which is characterized by sharp spectral features, often comparable to those routinely resolved at high resolution in the laboratory using supersonic jet sources.

\begin{table}
\centering
\footnotesize
\caption{\textbf{Laboratory measurements of hyperfine-split rotational transitions of ground state of benzonitrile.} Estimated experimental uncertainties (1$\sigma$) are in units of the last significant digit.  Calculated frequencies were derived from the best-fit constants listed under 'This Work' in Table~\ref{bn:constants}. }
\label{bn:lines}
\begin{tabular}{lrrr}
\hline\hline
\multicolumn{2}{c}{Transition}   & Frequency    & 	Obs.-Calc  \\
\cline{1-2}
  $J'_{K'_a,K'_c}\rightarrow J_{K_a,K_c}$  &  $F' \rightarrow F$  & (MHz)    & 	(MHz)  \\
\hline
$10_{2,8}\rightarrow10_{2,9}$  &      10$\rightarrow$10&      7004.0425(20)&      -0.0003\\
$10_{2,8}\rightarrow10_{2,9}$  &      11$\rightarrow$11&      7004.1529(20)&      -0.0020\\
$10_{2,8}\rightarrow10_{2,9}$  &      9$\rightarrow$9&      7004.1685(20)&      0.0024\\
$3_{0,3}\rightarrow2_{0,2}$  &      3$\rightarrow$3&      8205.4344(20)&      -0.0012\\
$3_{0,3}\rightarrow2_{0,2}$  &      2$\rightarrow$1&      8206.5649(20)&      0.0005\\
$3_{0,3}\rightarrow2_{0,2}$  &      3$\rightarrow$2&      8206.7917(20)&      0.0014\\
$3_{0,3}\rightarrow2_{0,2}$  &      4$\rightarrow$3&      8206.8303(20)&      0.0009\\
$3_{0,3}\rightarrow2_{0,2}$  &      2$\rightarrow$3&      8207.3158(20)&      -0.0017\\
$3_{0,3}\rightarrow2_{0,2}$  &      2$\rightarrow$2&      8208.6721(20)&      -0.0002\\
$7_{1,6}\rightarrow7_{1,7}$  &      7$\rightarrow$7&      9163.8456(20)&      0.0000\\
$7_{1,6}\rightarrow7_{1,7}$  &      8$\rightarrow$8&      9163.9557(20)&      0.0009\\
$7_{1,6}\rightarrow7_{1,7}$  &      6$\rightarrow$6&      9163.9699(20)&      -0.0007\\
$4_{0,4}\rightarrow3_{0,3}$  &      4$\rightarrow$4&      10853.8490(20)&      -0.0008\\
$4_{0,4}\rightarrow3_{0,3}$  &      3$\rightarrow$2&      10855.1359(20)&      0.0003\\
$4_{0,4}\rightarrow3_{0,3}$  &      4$\rightarrow$3&      10855.2443(20)&      0.0008\\
$4_{0,4}\rightarrow3_{0,3}$  &      5$\rightarrow$4&      10855.2618(20)&      0.0011\\
$4_{0,4}\rightarrow3_{0,3}$  &      3$\rightarrow$3&      10857.0171(20)&      -0.0004\\
$4_{3,2}\rightarrow3_{3,1}$  &      3$\rightarrow$3&      11080.1631(20)&      -0.0010\\
$4_{3,2}\rightarrow3_{3,1}$  &      4$\rightarrow$3&      11080.7925(50)&      0.0000\\
$4_{3,2}\rightarrow3_{3,1}$  &      5$\rightarrow$4&      11082.0566(50)&      0.0007\\
$4_{3,2}\rightarrow3_{3,1}$  &      3$\rightarrow$2&      11082.5444(50)&      -0.0003\\
$4_{3,1}\rightarrow3_{3,0}$  &      3$\rightarrow$3&      11084.3958(50)&      -0.0009\\
$4_{3,1}\rightarrow3_{3,0}$  &      4$\rightarrow$3&      11085.0238(50)&      0.0000\\
$4_{3,1}\rightarrow3_{3,0}$  &      5$\rightarrow$4&      11086.2876(50)&      -0.0004\\
$4_{3,1}\rightarrow3_{3,0}$  &      3$\rightarrow$2&      11086.7761(50)&      -0.0009\\
$5_{2,4}\rightarrow4_{2,3}$  &      5$\rightarrow$5&      13762.5220(20)&      -0.0010\\
$5_{2,4}\rightarrow4_{2,3}$  &      5$\rightarrow$4&      13763.0998(20)&      0.0008\\
$5_{2,4}\rightarrow4_{2,3}$  &      6$\rightarrow$5&      13763.3990(20)&      0.0006\\
$5_{2,4}\rightarrow4_{2,3}$  &      4$\rightarrow$3&      13763.4290(20)&      0.0005\\
$5_{2,4}\rightarrow4_{2,3}$  &      4$\rightarrow$4&      13764.1519(20)&      -0.0010\\
$5_{2,3}\rightarrow4_{2,2}$  &      5$\rightarrow$5&      14131.3238(20)&      -0.0010\\
$5_{2,3}\rightarrow4_{2,2}$  &      5$\rightarrow$4&      14131.9354(20)&      0.0009\\
$5_{2,3}\rightarrow4_{2,2}$  &      6$\rightarrow$5&      14132.2523(20)&      0.0013\\
$5_{2,3}\rightarrow4_{2,2}$  &      4$\rightarrow$3&      14132.2832(20)&      0.0003\\
$5_{2,3}\rightarrow4_{2,2}$  &      4$\rightarrow$4&      14133.0481(20)&      -0.0016\\
\hline\hline
\label{measurements}
\end{tabular}
\end{table}

\setcounter{table}{1}
\begin{table}[t]
\centering
\footnotesize
\caption{\textit{Continued} } 
\begin{tabular}{lrrr}
\hline\hline
\multicolumn{2}{c}{Transition}   & Frequency$^1$    & 	Obs.-Calc$^2$  \\
\cline{1-2}
  $J'_{K'_a,K'_c}\rightarrow J_{K_a,K_c}$  &  $F' \rightarrow F$  & (MHz)    & 	(MHz)  \\
\hline
$6_{1,6}\rightarrow5_{1,5}$  &      6$\rightarrow$6&      15432.6638(20)&      -0.0004\\
$6_{1,6}\rightarrow5_{1,5}$  &      6$\rightarrow$5&      15433.9164(20)&      0.0006\\
$6_{1,6}\rightarrow5_{1,5}$  &      5$\rightarrow$4&      15433.9335(20)&      0.0008\\
$6_{1,6}\rightarrow5_{1,5}$  &      7$\rightarrow$6&      15433.9687(20)&      0.0011\\
$6_{1,6}\rightarrow5_{1,5}$  &      5$\rightarrow$5&      15435.4388(20)&      -0.0007\\
$6_{2,5}\rightarrow5_{2,4}$  &      6$\rightarrow$6&      16479.6457(20)&      -0.0009\\
$6_{2,5}\rightarrow5_{2,4}$  &      6$\rightarrow$5&      16480.5224(20)&      0.0004\\
$6_{2,5}\rightarrow5_{2,4}$  &      5$\rightarrow$4&      16480.6985(100)$^*$&      0.0015\\
$6_{2,5}\rightarrow5_{2,4}$  &      7$\rightarrow$6&      16480.6985(100)$^*$&      0.0008\\
$6_{2,5}\rightarrow5_{2,4}$  &      5$\rightarrow$5&      16481.7500(20)&      -0.0009\\
$6_{1,5}\rightarrow5_{1,4}$  &      6$\rightarrow$6&      17388.3630(20)   &      0.0007\\
$6_{1,5}\rightarrow5_{1,4}$  &      6$\rightarrow$5&      17389.7305(20)   &      0.0002\\
$6_{1,5}\rightarrow5_{1,4}$  &      5$\rightarrow$4&      17389.7418(20)   &      0.0008\\
$6_{1,5}\rightarrow5_{1,4}$  &      7$\rightarrow$6&      17389.7815(20)   &      0.0012\\
$6_{1,5}\rightarrow5_{1,4}$  &      5$\rightarrow$5&      17391.3854(20)   &      0.0008\\
$7_{1,7}\rightarrow6_{1,6}$  &      7$\rightarrow$7&      17950.8538(20)   &      0.0007\\
$7_{1,7}\rightarrow6_{1,6}$  &      7$\rightarrow$6&      17952.1578(20)   &      0.0001\\
$7_{1,7}\rightarrow6_{1,6}$  &      6$\rightarrow$5&      17952.1662(20)   &      0.0027\\
$7_{1,7}\rightarrow6_{1,6}$  &      8$\rightarrow$7&      17952.1922(20)   &      0.0013\\
$7_{1,7}\rightarrow6_{1,6}$  &      6$\rightarrow$6&      17953.6871(20)   &      0.0001\\
$7_{0,7}\rightarrow6_{0,6}$  &      7$\rightarrow$7&      18407.9726(20)   &      0.0006\\
$7_{0,7}\rightarrow6_{0,6}$  &      6$\rightarrow$5&      18409.3497(20)   &      0.0006\\
$7_{0,7}\rightarrow6_{0,6}$  &      8$\rightarrow$7&      18409.3847(20)   &      0.0007\\
$7_{0,7}\rightarrow6_{0,6}$  &      6$\rightarrow$6&      18411.0010(20)   &      0.0018\\
$7_{2,6}\rightarrow7_{0,7}$  &      6$\rightarrow$6&      18993.8980(20)   &      0.0004\\
$7_{2,6}\rightarrow7_{0,7}$  &      8$\rightarrow$8&      18993.9347(20)   &      0.0008\\
$7_{2,6}\rightarrow7_{0,7}$  &      7$\rightarrow$7&      18994.1816(20)   &      0.0009\\
$7_{2,6}\rightarrow6_{2,5}$  &      7$\rightarrow$7&      19177.9504(20)   &      0.0003\\
$7_{2,6}\rightarrow6_{2,5}$  &      7$\rightarrow$6&      19178.9991(20)   &      0.0027\\
$7_{2,6}\rightarrow6_{2,5}$  &      6$\rightarrow$6&      19180.3305(20)   &      0.0012\\
$7_{6,2}\rightarrow6_{6,1}$  &      7$\rightarrow$6&      19382.0496(20)$^*$&      -0.0002\\
$7_{6,1}\rightarrow6_{6,0}$  &      7$\rightarrow$6&      19382.0496(20)$^*$&      -0.0004\\
$7_{6,2}\rightarrow6_{6,1}$  &      8$\rightarrow$7&      19382.9949(20)$^*$&      0.0010\\
$7_{6,1}\rightarrow6_{6,0}$  &      8$\rightarrow$7&      19382.9949(20)$^*$&      0.0009\\
$7_{6,2}\rightarrow6_{6,1}$  &      6$\rightarrow$5&      19383.1876(20)$^*$&      0.0002\\
$7_{6,1}\rightarrow6_{6,0}$  &      6$\rightarrow$5&      19383.1876(20)$^*$&      0.0000\\
\hline\hline
\end{tabular}
\end{table}

\setcounter{table}{1}
\begin{table}[t]
\centering
\footnotesize
\caption{\textit{Continued} } 
\begin{tabular}{lrrr}
\hline\hline
\multicolumn{2}{c}{Transition}   & Frequency$^1$    & 	Obs.-Calc$^2$  \\
\cline{1-2}
  $J'_{K'_a,K'_c}\rightarrow J_{K_a,K_c}$  &  $F' \rightarrow F$  & (MHz)    & 	(MHz)  \\
\hline
$7_{2,5}\rightarrow6_{2,4}$  &      7$\rightarrow$7&      20094.8441(20)&      -0.0012\\
$7_{2,5}\rightarrow6_{2,4}$  &      7$\rightarrow$6&      20095.9648(20)&      0.0002\\
$7_{2,5}\rightarrow6_{2,4}$  &      6$\rightarrow$5&      20096.0819(20)&      -0.0006\\
$7_{2,5}\rightarrow6_{2,4}$  &      8$\rightarrow$7&      20096.0931(20)&      0.0014\\
$7_{2,5}\rightarrow6_{2,4}$  &      6$\rightarrow$6&      20097.3891(20)&      -0.0018\\
$7_{1,6}\rightarrow6_{1,5}$  &      7$\rightarrow$7&      20191.5171(20)&      -0.0006\\
$7_{1,6}\rightarrow6_{1,5}$  &      6$\rightarrow$5&      20192.9347(100)$^*$&      0.0021\\
$7_{1,6}\rightarrow6_{1,5}$  &      7$\rightarrow$6&      20192.9347(100)$^*$&      0.0004\\
$7_{1,6}\rightarrow6_{1,5}$  &      8$\rightarrow$7&      20192.9641(20)&      0.0008\\
$7_{1,6}\rightarrow6_{1,5}$  &      6$\rightarrow$6&      20194.5885(20)&      -0.0001\\
$8_{1,8}\rightarrow7_{1,7}$  &      8$\rightarrow$8&      20451.5657(20)&      -0.0009\\
$8_{1,8}\rightarrow7_{1,7}$  &      7$\rightarrow$6&      20452.9046(100)$^*$&      0.0013\\
$8_{1,8}\rightarrow7_{1,7}$  &      8$\rightarrow$7&      20452.9046(100)$^*$&      0.0016\\
$8_{1,8}\rightarrow7_{1,7}$  &      9$\rightarrow$8&      20452.9267(20)&      0.0017\\
$8_{1,8}\rightarrow7_{1,7}$  &      7$\rightarrow$7&      20454.4322(20)&      -0.0004\\
$8_{0,8}\rightarrow7_{0,7}$  &      8$\rightarrow$8&      20826.7928(20)&      -0.0010\\
$8_{0,8}\rightarrow7_{0,7}$  &      7$\rightarrow$6&      20828.1760(20)&      0.0014\\
$8_{0,8}\rightarrow7_{0,7}$  &      9$\rightarrow$8&      20828.2022(20)&      0.0010\\
$8_{0,8}\rightarrow7_{0,7}$  &      8$\rightarrow$7&      20828.2065(20)&      0.0019\\
$8_{0,8}\rightarrow7_{0,7}$  &      7$\rightarrow$7&      20829.7878(20)&      -0.0017\\
$8_{2,7}\rightarrow7_{2,6}$  &      8$\rightarrow$8&      21854.7696(20)&      -0.0004\\
$8_{2,7}\rightarrow7_{2,6}$  &      8$\rightarrow$7&      21855.9330(20)&      0.0008\\
$8_{2,7}\rightarrow7_{2,6}$  &      7$\rightarrow$6&      21855.9943(20)&      -0.0002\\
$8_{2,7}\rightarrow7_{2,6}$  &      9$\rightarrow$8&      21856.0068(20)&      0.0004\\
$8_{2,7}\rightarrow7_{2,6}$  &      7$\rightarrow$7&      21857.3232(20)&      -0.0012\\
$9_{1,9}\rightarrow8_{1,8}$  &      9$\rightarrow$9&      22936.3739(20)&      -0.0003\\
$9_{1,9}\rightarrow8_{1,8}$  &      8$\rightarrow$7&      22937.7335(100)$^*$&      0.0031\\
$9_{1,9}\rightarrow8_{1,8}$  &      9$\rightarrow$8&      22937.7335(100)$^*$&      0.0008\\
$9_{1,9}\rightarrow8_{1,8}$  &      10$\rightarrow$9&      22937.7499(20)&      0.0020\\
$9_{1,9}\rightarrow8_{1,8}$  &      8$\rightarrow$8&      22939.2601(20)&      0.0001\\
$8_{1,7}\rightarrow7_{1,6}$  &      7$\rightarrow$6&      22943.4646(20)&      0.0006\\
$8_{1,7}\rightarrow7_{1,6}$  &      8$\rightarrow$7&      22943.4750(20)&      0.0020\\
$8_{1,7}\rightarrow7_{1,6}$  &      9$\rightarrow$8&      22943.4896(20)&      0.0011\\
$8_{1,7}\rightarrow7_{1,6}$  &      7$\rightarrow$7&      22945.1178(20)&      -0.0005\\
\hline\hline
\end{tabular}
\end{table}

\setcounter{table}{1}
\begin{table}[t]
\centering
\footnotesize
\caption{\textit{Continued} } 
\begin{tabular}{lrrr}
\hline\hline
\multicolumn{2}{c}{Transition}   & Frequency$^1$    & 	Obs.-Calc$^2$  \\
\cline{1-2}
  $J'_{K'_a,K'_c}\rightarrow J_{K_a,K_c}$  &  $F' \rightarrow F$  & (MHz)    & 	(MHz)  \\
\hline
$9_{0,9}\rightarrow8_{0,8}$  &      9$\rightarrow$9&      23226.3045(20)&      -0.0008\\
$9_{0,9}\rightarrow8_{0,8}$  &      8$\rightarrow$7&      23227.6869(20)&      -0.0034\\
$9_{0,9}\rightarrow8_{0,8}$  &      9$\rightarrow$8&      23227.7149(100)&      0.0021\\
$9_{0,9}\rightarrow8_{0,8}$  &      8$\rightarrow$8&      23229.2756(20)&      0.0004\\
$10_{1,10}\rightarrow9_{1,9}$  &      10$\rightarrow$10&      25407.4429(20)&      -0.0017\\
$10_{1,10}\rightarrow9_{1,9}$  &      9$\rightarrow$8&      25408.8173(100)$^*$&      0.0025\\
$10_{1,10}\rightarrow9_{1,9}$  &      10$\rightarrow$9&      25408.8173(100)$^*$&      -0.0009\\
$10_{1,10}\rightarrow9_{1,9}$  &      11$\rightarrow$10&      25408.8335(20)&      0.0043\\
$10_{1,10}\rightarrow9_{1,9}$  &      9$\rightarrow$9&      25410.3410(20)&      -0.0011\\
$9_{1,8}\rightarrow8_{1,7}$  &      9$\rightarrow$9&      25628.5875(20)&      -0.0012\\
$9_{1,8}\rightarrow8_{1,7}$  &      8$\rightarrow$7&      25630.0355(20)&      -0.0005\\
$9_{1,8}\rightarrow8_{1,7}$  &      9$\rightarrow$8&      25630.0497(20)&      -0.0001\\
$9_{1,8}\rightarrow8_{1,7}$  &      10$\rightarrow$9&      25630.0589(20)&      0.0029\\
$9_{1,8}\rightarrow8_{1,7}$  &      8$\rightarrow$8&      25631.6802(20)&      -0.0010\\
$9_{2,7}\rightarrow8_{2,6}$  &      9$\rightarrow$9&      26114.9381(20)&      -0.0005\\
$9_{2,7}\rightarrow8_{2,6}$  &      9$\rightarrow$8&      26116.2730(20)&      0.0009\\
$9_{2,7}\rightarrow8_{2,6}$  &      8$\rightarrow$7&      26116.3190(20)&      -0.0016\\
$9_{2,7}\rightarrow8_{2,6}$  &      10$\rightarrow$9&      26116.3340(20)&      0.0013\\
$9_{2,7}\rightarrow8_{2,6}$  &      8$\rightarrow$8&      26117.8218(20)&      -0.0004\\
$10_{2,9}\rightarrow9_{2,8}$  &      10$\rightarrow$10&      27135.7430(20)&      -0.0020\\
$10_{2,9}\rightarrow9_{2,8}$  &      10$\rightarrow$9&      27137.0352(20)&      0.0025\\
$10_{2,9}\rightarrow9_{2,8}$  &      9$\rightarrow$8&      27137.0583(20)&      0.0002\\
$10_{2,9}\rightarrow9_{2,8}$  &      11$\rightarrow$10&      27137.0719(20)&      0.0029\\
$10_{2,9}\rightarrow9_{2,8}$  &      9$\rightarrow$9&      27138.4890(20)&      -0.0009\\
$11_{1,11}\rightarrow10_{1,10}$  &      11$\rightarrow$11&      27867.2017(20)&      0.0006\\
$11_{1,11}\rightarrow10_{1,10}$  &      10$\rightarrow$9&      27868.5833(100)$^*$&      0.0014\\
$11_{1,11}\rightarrow10_{1,10}$  &      11$\rightarrow$10&      27868.5833(100)$^*$&      -0.0024\\
$11_{1,11}\rightarrow10_{1,10}$  &      12$\rightarrow$11&      27868.5984(20)&      0.0045\\
$11_{1,11}\rightarrow10_{1,10}$  &      10$\rightarrow$10&      27870.1061(20)&      0.0004\\
$11_{0,11}\rightarrow10_{0,10}$  &      11$\rightarrow$11&      28018.6086(20)&      -0.0016\\
$11_{0,11}\rightarrow10_{0,10}$  &      10$\rightarrow$9&      28019.9998(20)&      -0.0026\\
$11_{0,11}\rightarrow10_{0,10}$  &      12$\rightarrow$11&      28020.0150(20)&      -0.0003\\
$11_{0,11}\rightarrow10_{0,10}$  &      10$\rightarrow$10&      28021.5459(20)&      -0.0023\\
$10_{1,9}\rightarrow9_{1,8}$  &      10$\rightarrow$10&      28242.2717(20)&      -0.0008\\
$10_{1,9}\rightarrow9_{1,8}$  &      9$\rightarrow$8&      28243.7213(20)&      -0.0013\\
$10_{1,9}\rightarrow9_{1,8}$  &      11$\rightarrow$10&      28243.7419(100)$^*$&      0.0026\\
$10_{1,9}\rightarrow9_{1,8}$  &      10$\rightarrow$9&      28243.7419(100)$^*$&      0.0020\\
$10_{1,9}\rightarrow9_{1,8}$  &      9$\rightarrow$9&      28245.3536(20)&      -0.0004\\
$10_{3,8}\rightarrow10_{1,9}$  &      9$\rightarrow$9&      28749.9255(20)&      0.0000\\
$10_{3,8}\rightarrow10_{1,9}$  &      11$\rightarrow$11&      28749.9571(20)&      0.0008\\
$10_{3,8}\rightarrow10_{1,9}$  &      10$\rightarrow$10&      28750.2628(20)&      -0.0008\\
\hline\hline
\end{tabular}
\\
$^{*}$ Blended hyperfine feature. \\ 
\end{table}

\clearpage

The best-fitting spectroscopic constants for benzonitrile are reported in Table~\ref{bn:constants}.  $A,B,$ and $C$ are rotational constants, $\Delta_J$, $\Delta_{JK}$, $\Delta_K$, $\delta_J$, and $\delta_K$ are centrifugal distortion constants, and $\chi_{aa}$(N) and $\chi_{bb}$(N) are nuclear hyperfine coupling constants.

\begin{table}[ht!]
\centering
\caption{\textbf{Benzonitrile spectroscopic constants.} Best-fitting spectroscopic constants for the ground state of benzonitrile derived from the laboratory measurements in Table~\ref{bn:lines}.  Uncertainties (1$\sigma$) are given in parentheses are in units of the last significant digit.  The global constants are derived from a combination of the low-frequency FT microwave measurements reported in \cite{Wohlfart:2008hg} and those in Table~\ref{bn:lines}.}
\label{bn:constants}
\begin{tabular}{lrrr}
\hline\hline
Constant        	  &  Ref.~\cite{Wohlfart:2008hg} & This work$^{1,2}$    & 	Global$^{1,3}$  \\
\hline
$A$ (MHz)            &	5655.2654(72)	    &	5655.26522(59)    		&   5655.26519(59) 		  \\
$B$ (MHz)            &	1546.875864(66)	    &	1546.875836(63)   		&	1546.875822(54) 		  \\
$C$ (MHz)            &	1214.40399(10)	    &	1214.404061(48)   		&	1214.404047(40) 		  \\
\\					
$\Delta_J$ (kHz)     &	0.0456(15)		    &	0.045629(284)   		&	0.04555(235) 		  \\
$\Delta_{JK}$ (kHz)  &	0.9381(56)		    &	0.93328(241)   			&	0.93304(234) 		  \\
$\Delta_K$ (kHz)     &	0.50(38)		    &	0.272(64)   			&	0.272(64) 		  \\
$\delta_J$ (kHz)     &	0.01095(41)		    &	0.011106(163)   		&	0.011094(157) 		  \\
$\delta_K$ (kHz)     &	0.628(53)		    &	0.6136(73)   			&	0.6141(72) 		  \\
\\		
$\chi_{aa}$(N) (MHz) &	-4.23738(36)        &	-4.23797(89)   			&	-4.23749(45) 		  \\
$\chi_{bb}$(N) (MHz) &	 2.2886(11)	        &     2.28907(118)   		&	 2.28871(65) 		  \\
\\
Number of Measurements &      78            &     146                   &      175 \\
$\sigma$ (MHz)       &   0.00524            &    0.00130                &    0.00123  \\
weighted average     &   0.709              &     0.622                 &     0.628  \\
\hline\hline
\end{tabular}
\end{table}

\begin{table}
\centering
\caption{\textbf{Constants comparisons.}  Comparison of best-fitting spectroscopic constants and associated uncertainties of this work relative to that of ~\cite{Wohlfart:2008hg}.  $\Delta$ is the difference in the best-fit value from \cite{Wohlfart:2008hg} compared to this work, divided by its uncertainty ($\sigma$) from\cite{Wohlfart:2008hg}.  The final column is the ratio of the uncertainty ($\sigma$) from \cite{Wohlfart:2008hg} relative to that derived from this work (Table~\ref{bn:constants}). Ratios greater than 1 indicate improved precision for that constant in the new study.}
\label{bn:compare1}
\begin{tabular}{lrr}
\hline\hline
Constant	& $\Delta/\sigma$~$^1$   &  Ratio of $\sigma$~$^2$     \\
\hline
$A$           &	0.03	&	12.20  \\
$B$           &	0.42	&	1.05   \\
$C$           &	-0.71	&	2.08   \\
\\					
$\Delta_J$    &	-0.02	&	5.28  \\
$\Delta_{JK}$ &	0.86	&	2.32    \\
$\Delta_K$    &	0.60	&	5.94  \\
$\delta_J$    &	-0.38	&	2.52    \\
$\delta_K$    &	0.27	&	7.26    \\
\\					
$\chi_{aa}$(N) &	1.64	&	0.40   \\
$\chi_{bb}$(N) &	-0.43	&	0.93    \\
\hline
\end{tabular}
\end{table}

\begin{table}
\centering
\caption{\textbf{Global fit constants comparison.} Same as Table~\ref{bn:compare1}, but comparing the global fit to just this work.}
\label{bn:compare2}
\begin{tabular}{lrr}
\hline\hline
Constant	& $\Delta/\sigma$~$^1$   &  Ratio of $\sigma$~$^2$     \\
\hline
$A$          &	 0.05	&   1.00  \\
$B$          & 	0.22	&   1.17  \\
$C$          & 	0.29	&   1.20	  \\
\\					
$\Delta_J$    & 	0.28	&   1.21  \\
$\Delta_{JK}$ & 	0.10	&   1.03  \\
$\Delta_K$    & 	0.00	&   1.00  \\
$\delta_J$    &	0.07	&   1.04  \\
$\delta_K$    &	-0.07	&   1.01  \\
\\					
$\chi_{aa}$(N) &  -0.54	&   1.98  \\
$\chi_{bb}$(N) &   0.31	&   1.82  \\
\hline
\end{tabular}
\end{table}


\end{document}